\documentclass[fdp,fleqn]{w-art}
\usepackage{times}
\usepackage{w-thm}
\usepackage[]{graphicx}
\begin{document}
\pagespan{1}{}
{\small KUL-TF-08/29;$\,$ ITP-UU-08/66;$\,$ SPIN-08/53}



\title[Screening on the conifold]{Non chiral dynamical flavors and screening on the conifold}


\author[F. Bigazzi]{Francesco Bigazzi\inst{1}\footnote{E-mail:~\textsf{fbigazzi@ulb.ac.be}}}
\address[\inst{1}]{Physique Th\'eorique et Math\'ematique and International Solvay
Institutes, Universit\'e Libre de Bruxelles; CP 231, B-1050
Bruxelles, Belgium.}
\author[A. L. Cotrone]{Aldo L. Cotrone\inst{2}\footnote{E-mail:~\textsf{Aldo.Cotrone@fys.kuleuven.be}}}
\address[\inst{2}]{Institute for theoretical physics, K.U. Leuven;
Celestijnenlaan 200D, B-3001 Leuven,
Belgium.}
\author[A. Paredes]{Angel Paredes\inst{3}\footnote{E-mail:~\textsf{A.ParedesGalan@uu.nl}}}
\address[\inst{3}]{Institute for Theoretical Physics, Utrecht University; Leuvenlaan 4,
3584 CE Utrecht, The Netherlands.}
\author[A. Ramallo]{Alfonso Ramallo\inst{4}\footnote{E-mail:~\textsf{alfonso@fpaxp1.usc.es}}}
\address[\inst{4}]{Departamento de Fisica de Particulas, Universidade de Santiago de Compostela and
Instituto Galego de Fisica de Altas Energias (IGFAE); E-15782, Santiago de Compostela, Spain.}

\begin{abstract}
We present a new class of string theory solutions which are conjectured to be dual to the ${\cal N}=1$ conifold theory by Klebanov and Witten coupled to non chiral massive dynamical flavors. These are introduced, in the Veneziano limit, by means of suitably embedded and smeared D7-branes whose full backreaction on the background is taken into account. The string solutions are used to study chromoelectric charge screening effects due to the dynamical flavors in the non perturbative regime of the dual gauge theories.
\end{abstract}
\maketitle                   





\section{Introduction}
The nonperturbative dynamics of QCD is crucially affected by the contribution of the light quarks. One of the most striking effects is the maximal screening of the chromoelectric charges due to fundamental pair production and chromoelectric flux breaking. On the difficult road towards constructing a string dual of (planar) QCD or its ${\cal N}=1$ supersymmetric versions, it is thus mandatory to take care of the fundamental fields. These are introduced by means of $N_f$ ``flavor'' D-brane sources on a closed string background arising from the backreaction of both the ``flavor'' and $N_c\gg1$ ``color'' D-branes. The flavor branes share with the color ones the (4d) Minkowski directions and wrap transverse non compact manifolds: this is how the global flavor symmetry is implemented. If the flavor branes are on top of each other this symmetry is non abelian. This obviously interesting setup is quite hard to deal with in practice if the flavor brane backreaction is not neglected: the supergravity equations of motion are a set of coupled partial differential equations with delta function sources. An enormous simplification can be obtained if the number of flavors is very large (as in the Veneziano limit $N_c,N_f\rightarrow\infty$ with $N_f/N_c$ and $g^2 N_c$ fixed): by homogeneously smearing the flavor branes in the transverse space, delta function sources are replaced by density distributions and the relevant equations to solve are ordinary differential equations in a radial variable \cite{paris}. The price to pay is the generic breaking of the flavor symmetry group into a product of abelian factors. 

If the flavors are massless, the flavor branes reach the origin of the transverse space. In the smeared setup the origin is a very special point: since all the $N_f$ branes overlap there, the flavor symmetry group is generically enhanced. A global symmetry enhancement is usually signaled by a singularity in the string background: indeed all the known smeared string solutions dual to massless-flavored theories have a (good) singularity at the origin. If the flavors are massive the corresponding D-branes extend up to a certain distance from the origin and in the smeared setups there is no special point where the flavor symmetry is fully enhanced: the singularity due to the flavors disappears and is replaced by a region where there is no flavor D-brane charge at all.\footnote{The setup is sensible provided tadpole cancellation is guaranteed.} The smeared density distribution, then,
depends on the radial coordinate and stops at the boundary of that region. As a result, the string solution inside that region is
just an unflavored one: this is dual to the fact that for energies smaller than the flavor mass the flavors are integrated out.

In this paper we find new explicit string solutions accounting for the full backreaction of $N_c$ D3 and $N_f$ smeared D7-branes
on the conifold. Each D7-brane is holomorphically embedded in a particular way \cite{ouyang,kuper} that makes the corresponding flavor multiplets massive and the theory ``non chiral'' (the massless ``chiral'' setup was examined in \cite{Benini:2006hh} and the massive one in \cite{massivekw}). The unflavored gauge theory is the ${\cal N}=1$ conformal $SU(N_c)\times SU(N_c)$ Klebanov-Witten (KW) quiver theory \cite{kw}. Each D7-brane  adds a (anti) fundamental flavor to a node of the gauge group. The resulting gauge theory has a UV Landau pole and a superpotential containing only quartic couplings and flavor mass terms. 

The new string solutions are then used to get quantitative predictions on the screening effects on the static quark-antiquark potential in the dual field theory at strong coupling. The analysis shows that both the chromoelectric force between static sources and the screening length are reduced as the number of flavors (resp. their mass) is increased (resp. reduced).

\section{Non chiral embeddings on the conifold}
The singular conifold is a six dimensional Calabi-Yau cone whose defining equation in $\mathbf C^4$ can be written as $z_1\,z_2 - z_3\,z_4 = 0$. The base of the conifold is the $T^{1,1}$ Sasaki-Einstein manifold whose isometry group is $SU(2)\times SU(2)\times U(1)$. The conifold metric can be written as $ds^2_{C}= dr^2+r^2ds^2_{T^{1,1}}$, where
\begin{equation}
ds^2_{T^{1,1}} = \frac{1}{6}\sum_{i=1}^2[d\theta_i^2+ \sin^2\theta_id\varphi_i^2] + \frac{1}{9}[d\psi+\sum_{i=1}^2\cos\theta_id\varphi_i]^2\,,
\end{equation}
and $\psi \in [0,4\pi)$, $\varphi_i \in [0,2\pi)$, $\theta_i \in [0,\pi]$. The complex variables $z_i$ can be expressed in terms of the above coordinates. 

The low energy dynamics of $N_c$ D3-branes stuck at the tip of the conifold is described by the Klebanov-Witten (KW) theory \cite{kw}. This is an ${\cal N}=1$ superconformal quiver theory
with gauge group $SU(N_c)\times SU(N_c)$ and bifundamental matter fields $A, B$, arranged in $SU(2)\times SU(2)$ doublets, interacting through a quartic superpotential $W_{KW}=\epsilon^{ij}\epsilon^{kl}A_iB_kA_jB_l$. The $SU(2)\times SU(2)\times U(1)_R$ symmetric IR fixed point of the theory is conjectured to have a dual description in terms of type IIB string theory on an $AdS_5\times T^{1,1}$ background with $N_c$ units of five-form flux through $T^{1,1}$ and constant dilaton.

Let us now consider a particular and well studied way of adding fundamental degrees of freedom to the model, i.e. by means of $\kappa$-symmetric D7-branes wrapping the holomorphic 4-cycle defined by an equation like $z_1-z_2=\mu$ \cite{ouyang,kuper}. We call this embedding ``non chiral'' because, on the field theory side, a D7-brane adds a fundamental and an antifundamental flavor to one node of the KW quiver and the flavor mass terms do not break the classical flavor symmetry of the massless theory. The related superpotential is \cite{ouyang}
\begin{equation}
W = W_{KW} +\hat h_1\, \tilde q_1 [A_1B_1-A_2B_2 ]q_1 + \hat h_2\,\tilde q_2 [B_1A_1-B_2A_2]q_2 + k_i\,(\tilde q_i q_i)^2 + m\,(\tilde q_i q_i)\,,
\end{equation}
where we have considered a $\mathbf Z_2$-invariant setup where there are two stacks of D7 branes adding the same number of fundamental degrees of freedom to both nodes. Apart from the mass terms, the superpotential contains only quartic couplings.\footnote{In the case of "chiral" embeddings like $z_1=\mu$, the flavors induce cubic couplings in $W$.} The embedding preserves a $SU(2)$ diagonal subgroup of the conifold symmetries. Moreover both the scale invariance and the $U(1)$ isometry of the conifold are explicitly broken when $\mu\neq 0$. This is precisely mapped to the explicit breaking of conformal invariance and $U(1)_R$ symmetry driven by the flavor masses (which are then related to the $\mu$ parameter). In the massless case ($\mu=0$) the previous symmetries are broken at the quantum level. 

Let us now consider acting with an $SO(4)\sim SU(2)\times SU(2)$ rotation on the embedding equation $z_1-z_2=\mu$. This way we obtain the generalized embedding 
\begin{equation}
\bar p\,z_1 - p\,z_2 + \bar q\,z_3 + q\,z_4 = \mu\,,
\label{genz}
\end{equation}
where $p, q$ span a unit 3-sphere
\begin{equation}
p=\cos\frac{\theta}{2}e^{i(\frac{\chi+\phi}{2})}\,,\quad q=\sin\frac{\theta}{2}e^{i(\frac{\chi-\phi}{2})}\,,
\end{equation}
and $\chi \in [0,4\pi)$, $\phi \in [0,2\pi)$, $\theta \in [0,\pi]$. We can rewrite (\ref{genz}) more explicitly as
\begin{equation}
e^{\frac{3}{2}\rho} e^{-\frac{i}{2}(\chi+\phi)}e^{\frac{i}{2}(\psi-\varphi_1-\varphi_2)}\left[\Gamma_1 + \Gamma_2\right] = e^{\frac{3}{2}\rho_q}e^{i\beta}\,,
\label{genang}
\end{equation}
where we have defined (for reasons which will become clear in the following section) $r=e^{\rho}$, $e^{\frac{3}{2}\rho_q}e^{i\beta}\equiv\mu$ ($\beta\in[0,2\pi]$) and
\begin{eqnarray}
\Gamma_1 &=& \sin\frac{\theta}{2}\left[e^{i(\chi + \varphi_2)} \cos\frac{\theta_2}{2}\sin\frac{\theta_1}{2} + e^{i(\phi + \varphi_1)} \cos\frac{\theta_1}{2}\sin\frac{\theta_2}{2}\right]\,, \nonumber \\
\Gamma_2 &=& \cos\frac{\theta}{2}\left[\sin\frac{\theta_1}{2}\sin\frac{\theta_2}{2} - e^{i(\chi+\phi + \varphi_1+\varphi_2)} \cos\frac{\theta_1}{2}\cos\frac{\theta_2}{2}\right]\,.
\end{eqnarray}
We can split the complex equation (\ref{genang}) in two conditions 
\begin{eqnarray}
f_1 &\equiv& -\chi -\phi + \psi -\varphi_1 - \varphi_2 + 2\arg(\Gamma_1 + \Gamma_2) -2\beta +4\pi n=0\,,\nonumber\\
f_2 &\equiv& e^{3\rho} |\Gamma_1 +\Gamma_2|^2 - e^{3\rho_q} = 0 \,,
\end{eqnarray}
following from the phase and the (squared) modulus, respectively. These equations imply that the phase of the mass terms, related to $\beta$, drives the breaking of the shift symmetry along the $\psi$ direction (the dual of the $U(1)_R$ symmetry). Moreover, since $|\Gamma_1 +\Gamma_2|^2\le 1$, the D7 branes are extended along the radial direction only up to $\rho=\rho_q$. The energy of a string stretched along $\rho$ from the origin $\rho=-\infty$ to $\rho_q$ is a parameter $m_q$ which is conventionally called ``constituent mass'' of the flavors. When $\rho_q\rightarrow -\infty$, $m_q\rightarrow 0$.
\subsection{Smearing the embeddings}
\label{smecon}
Let us now consider $N_f \gg 1$ branes suitably distributed
within the family of embeddings (\ref{genang}).
In particular, we want to restore effectively the $SU(2)\times SU(2)
\times U(1)$ symmetry of $T^{1,1}$ which is partially broken by flavor branes on top of each other. Accordingly, we place branes at different values of
$\chi, \phi,
\theta, \beta$ with a suitably normalized symmetry preserving distribution $32\pi^3\rho_{\theta,\phi,\chi,\beta} = N_f \sin\theta$.

The D7 branes are sources of an $F_1$ RR field with modified Bianchi identity $dF_1 =- g_s \Omega$, where
\begin{equation}
\Omega = \int \rho_{\theta,\phi,\chi,\beta} \left(\delta(f_1) \delta(f_2) df_1 \wedge df_2
\right) d\theta\, d\phi\, d\chi\, d\beta\,,
\label{generalOmega}
\end{equation}
is the density distribution form of the smeared D7-branes. The general form of $\Omega$ is dictated by the symmetry, the only possibility for an exact two-form preserving $SU(2) \times SU(2) \times U(1)_\psi \times Z_2$ being \cite{Benini:2006hh}
\begin{equation}
\Omega = \frac{N_f(\rho)}{4\pi}(\sin\theta_1 d\theta_1 \wedge d\varphi_1
+ \sin\theta_2 d\theta_2 \wedge d\varphi_2)
-\frac{\dot N_f(\rho)}{4\pi} d\rho\wedge(d\psi + \cos \theta_1 d\varphi_1 + 
\cos \theta_2 d\varphi_2)\,,
\label{massive_2form}
\end{equation}
where the function $N_f(\rho)$ has to follow from the smeared embedding and the dot means derivative w.r.t. $\rho$. Notice, from eq. (\ref{massive_2form}), that $\Omega_{\rho\psi}$ is expected to be independent on $\theta_i,\varphi_i$ and so we can just pick, say, $\theta_1=\theta_2=0$ in the corresponding expression deduced from (\ref{generalOmega}). The integral is then easy to solve and the final result is $ 4\pi\Omega_{\rho\psi} =  - 3 N_f \Theta(\rho-\rho_q) e^{3(\rho_q-\rho)}$ which, together with (\ref{massive_2form}), gives $\dot N_f(\rho) = 3N_f \Theta(\rho-\rho_q) e^{3(\rho_q-\rho)}$. The ``running effective number of flavors'' thus reads
\begin{equation}
N_f(\rho)= 0 \,, \quad (\rho\le\rho_q)\,;\qquad N_f(\rho)= N_f \left(1-e^{3(\rho_q-\rho)}\right)\,, \quad (\rho>\rho_q)\,,
\label{nfcon}
\end{equation}
where the integration constants have been fixed by requiring continuity at $\rho_q$.\footnote{In the chiral case $\,N_f(\rho)=N_f [1 - e^{3\rho_q-3\rho} ( 1 +3\rho - 3\rho_q)]\,$ for $\rho>\rho_q\,$ \cite{massivekw}.} As we have anticipated in the introduction, there is an effectively unflavored region $\rho\le\rho_q$ where the D7 density distribution vanishes. In the dual gauge theory, this corresponds to the $E\le m_q$ regime where the flavors are integrated out. Moreover, $N_f(\rho\rightarrow\infty)=N_f$ and $N_f(\rho_q\rightarrow-\infty)=N_f$. The function $N_f(\rho)$ asymptotically approaches the constant value $N_f$, it has the shape of a smoothed
out Heaviside function, and it is actually well approximated by $N_f\Theta(\rho-\rho_q)$ in the small mass limit $\rho_q\rightarrow-\infty$.\footnote{The step function approximation was used in a SQCD-like setup with wrapped D5-branes \cite{Bigazzi:2008gd} to construct the massive versions of the unquenched solutions found in \cite{cnp}.} This shape reflects the fact that at energies $E\gg m_q$ the gauge theory effectively behaves as the massless-flavored one.
\section{The backreacted solution}
The ansatz for the (Einstein frame) metric of the backreacted flavored KW background is of the form \cite{Benini:2006hh}
\begin{eqnarray}
ds^2 &=& h^{-\frac12}(\rho) dx_{1,3}^2 + h^\frac12(\rho) ds_6^2\,, \nonumber \\
ds_6^2&=&
e^{2 f(\rho)} d\rho^2 + \frac{e^{2g(\rho)}}{6}\sum_{i=1,2}
(d\theta_i^2 + \sin^2 \theta_i d\varphi_i^2)
+\frac{e^{2f(\rho)}}{9}(d\psi + \sum_{i=1,2} \cos \theta_i d\varphi_i)^2\,,
\end{eqnarray}
which is a deformation of the KW metric driven by the functions  $f(\rho), g(\rho)$. The KW solution is obtained when $f(\rho)=g(\rho)=\rho$. The $\kappa$-symmetry equations for D7 branes on this deformed background can be obtained from those on the KW solution, just by substituting the KW radial coordinate $r$ with $e^{\rho}$.
The ansatz for the dilaton and the RR forms reads
\begin{eqnarray}
\phi&=&\phi(\rho)\,,\qquad F_5 = d\,h^{-1}(\rho)\wedge dx^0\wedge\dots\wedge dx^3 + {\rm Hodge\,dual}\,, \nonumber \\
F_1 &=& \frac{g_sN_f(\rho)}{4\pi} \bigl( d\psi + \cos\theta_1\,
d\varphi_1 + \cos\theta_2\, d\varphi_2 \bigr)\,\,.
\end{eqnarray}
The latter form explicitly solves the modified Bianchi identity $dF_1=-g_s\Omega$. Following a general strategy suggested by \cite{km} we look for a solution of the IIB supergravity plus (smeared) D7-brane DBI+WZ action. As discussed in \cite{massivekw,carlos} the use of the above D-brane action is well justified in the smeared setups, since the large transverse distance between two generic branes effectively suppresses the coupling $g_s N_f$. Along the same lines of \cite{Benini:2006hh}, the solution can be found from the supersymmetric variations of the bulk fermions giving rise to the following set of first order differential equations 
\begin{eqnarray}
&&\dot g\,=\,e^{2f-2g}\,, \qquad \dot f\,=\,3\,-\,2 e^{2 f-2 g}\,-\,\frac{3 g_s N_f(\rho)}{8\pi}\,e^{\phi}\,, \nonumber\\
&& \dot \phi\,=\,\frac{3 g_s N_f(\rho)}{4\pi}\,e^{\phi}\,,\qquad \dot h\,=\,- 27 \pi g_s N_c \alpha'^2\, e^{-4g}\,\,.
\end{eqnarray}
These have exactly the same form as those obtained in the (chiral) massless case by  \cite{Benini:2006hh}, modulo the substitution of $N_f$ with $N_f(\rho)$: this is due to the fact that the supersymmetric fermionic variations only contain the forms $F$ and not the $dF$ terms. 

The expression for the dilaton can be determined by direct integration. One gets, for $\rho>\rho_q$
\begin{equation}
e^{\phi}\,=\,-\,\,\frac{4\pi}{g_s N_f\,\big[\,3\rho\,-\,e^{3\rho_q-3\rho}\,\big(\,e^{3\rho}-1\,\big)\,\big]}\,\,,
\label{dilaton}
\end{equation}
where the integration constant has been adjusted to have dilaton divergence at $\rho=0$. This point is mapped to the field theory UV Landau pole. When $\rho\le\rho_q$, $e^{\phi}=e^{\phi_{IR}}\equiv e^{\phi(\rho_q)}$ by continuity.

Along the same lines, we can solve the remaining equations separately for $\rho < \rho_q$
and $\rho > \rho_q$. Then we demand continuity at $\rho = \rho_q$, regularity at $\rho\rightarrow -\infty$ and, as in  \cite{massivekw}, we fix the remaining integration constants so that $h(0)=g(0)=0$. The result is the following analytic solution.

For $\rho\in(-\infty,\rho_q]$ we have
\begin{eqnarray}
e^g &=& e^f = a \, e^{\rho}\,,\quad a=[2e^{3\rho_q}-2-6\rho_q]^{\frac16}[1-e^{3\rho_q}]^{-\frac13}\,, \nonumber \\
h(\rho)&=& 27 \pi g_s N_c \alpha'^2(b + \frac{e^{-4\rho}-e^{-4\rho_q}}{4a^4})\,,\quad  b=\int_{\rho_q}^0e^{-4g(\rho_*)}d\rho_*\,. 
\end{eqnarray}
This is just a slight deformation (driven by the constant factors) of the $AdS_5\times T^{1,1}$ background, as can be realized
changing variable to $r=e^{\rho}$. 
In the $\rho\in(\rho_q,0]$ region we have 
\begin{eqnarray}
&&e^{f}\,=\,[1-e^{3\rho_q}]^{-\frac13}\,\,
\frac{\Big[\,-6\rho e^{6\rho}\,+\,2 e^{3\rho+3\rho_q}\,\big(\,e^{3\rho}-1\,\big)\,\Big]^{\frac12}}
{\Big[\,\big(1-6\rho)e^{6\rho}\,+\,2 e^{3\rho+3\rho_q}\,\big(\,e^{3\rho}-2\,\big)\,+\,
e^{6\rho_q}\,\Big]^{{\frac13}}}\,\,,\nonumber \\
&&e^{g}\,=\,[1-e^{3\rho_q}]^{-\frac13}\,\,
 \Big[\,\big(1-6\rho)e^{6\rho}\,+\,2 e^{3\rho+3\rho_q}\,\big(\,e^{3\rho}-2\,\big)\,+\,
e^{6\rho_q}\,\Big]^{{\frac16}}\,\,,\nonumber \\
&&h(\rho)= 27 \pi g_s N_c \alpha'^2 \int_\rho^{0} e^{-4g(\rho_*)} d\rho_*\,\,.
\end{eqnarray}
It can be checked that for $\rho<0$ the above solution has small curvature provided $\,1\ll N_f\ll N_c$. This condition also ensures that the distance between two generic flavor branes is large (in string units).
\section{Screening from holography}
The background we have just found can be used to study how the nonperturbative gauge theory dynamics is affected by the dynamical flavors. The gauge theory we are talking about, the KW quiver coupled to massive non chiral flavors,
has 2 nodes with $N_c$ colors and $N_f+2N_c$ flavors each. The two gauge couplings run in the same direction (the beta functions have the same positive signs) and the net result is a QED-like theory with a UV Landau pole at a scale $\Lambda_{UV}$. 
It is interesting to ask how the expected Coulomb-like interaction between fundamental chromoelectric charges is affected by the sea flavors. To answer this question we can probe the theory with an external quark-antiquark pair $\bar Q, Q$ with mass $M_Q\gg m_q$, ($M_Q<\Lambda_{UV}$). We can then study how the static quark-antiquark potential depends on the sea quark parameters $m_q, N_f$ using well known holographic maps \cite{maldawilson}.
The $\bar QQ$ bound state is dual to an open string attached to a probe flavor-brane which extends up to $\rho_Q\gg \rho_q$, ($\rho_Q<0$). The string bends in the bulk up to a minimal radial position $\rho_0$.
The open string embedding is chosen as $t=\tau, y=\sigma, \rho=\rho(y)$ where $y\in [-L/2,L/2]$ is in Minkowski. The quark-anti quark distance $L$ as well as the renormalized (by mass subtraction) potential $V$ are then given by
\begin{equation}
L(\rho_0)=2\int_{\rho_0}^{\rho_{Q}} \frac{G F_{0}}{F\sqrt{F^2
-F_{0}^2}}d\rho\,,\quad  V (\rho_0)=\frac{2}{2\pi\alpha'} \Bigl[\int_{\rho_0}^{\rho_{Q}}
\frac{G F}{\sqrt{F^2 -F_{0}^2}}d\rho -
\int_{-\infty}^{\rho_{Q}} G \ d\rho \Bigr]\,,
\end{equation}
where $F\equiv\sqrt{g_{tt}g_{yy}},\,G=\sqrt{g_{tt}g_{\rho\rho}}\,$ and $F_0=F(\rho_0)$.

\begin{figure}
\centering
\includegraphics[width=0.3\textwidth]{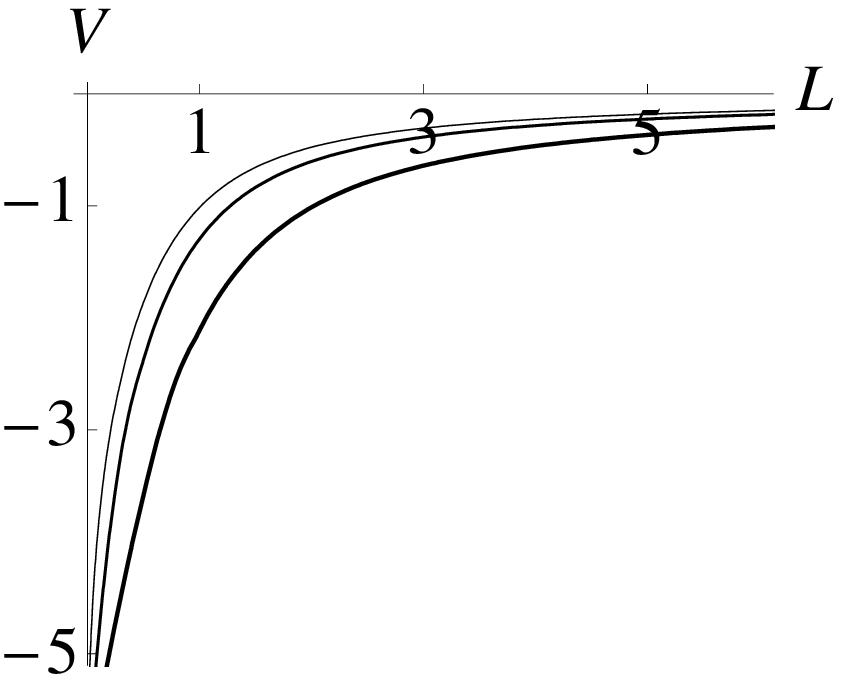}
\includegraphics[width=0.3\textwidth]{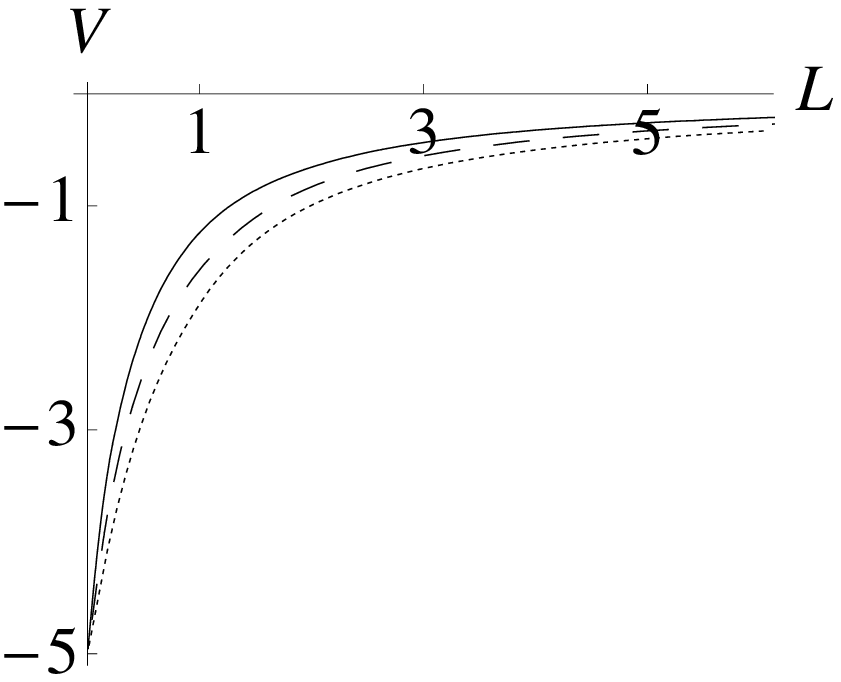}
\includegraphics[width=0.3\textwidth]{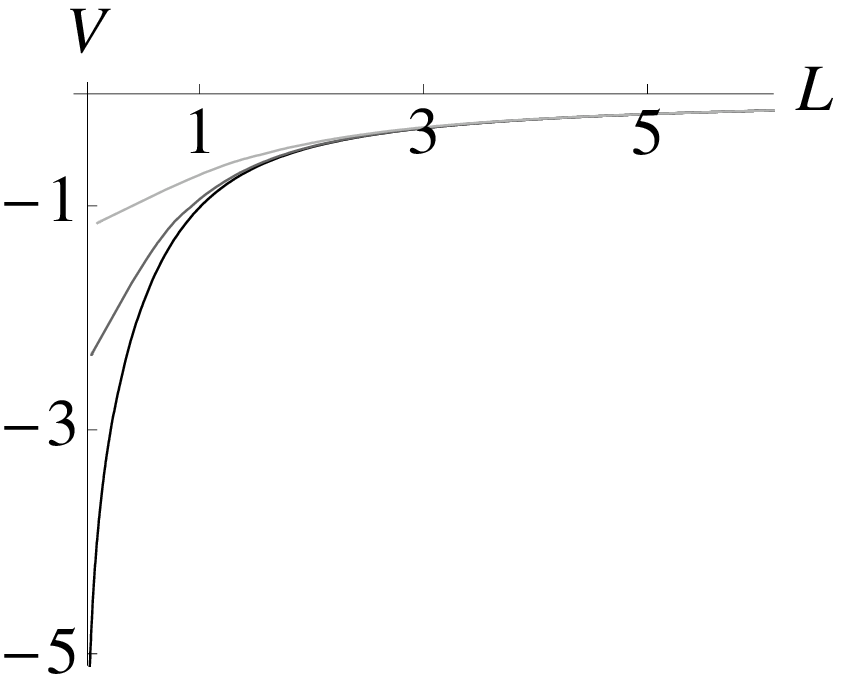}
\caption{From left to right, plots of $V(L)$ at different values of: $g_s
N_f=1, 0.6, 0.2$, top to bottom (at $m_q=0.1$, $M_Q=3$); $m_q=0.5, 1, 1.5$, top to bottom (at $g_s N_f=1$, $M_Q=2.5$); $M_Q=0.6, 1.2, 3$, top to bottom (at $g_sN_f=1$, $m_q=0.1$). We have put $(2\pi\alpha')^{-1}=1$ and $\sqrt{27\pi g_s N_c \alpha'^2}=1$ in the expressions for $V$ and $L$ respectively.}
\label{V}
\end{figure}

Using the explicit backreacted solution found above we find (see figure \ref{V}) that the static quark-antiquark potential is negative and has the same Coulomb-like behavior as the unflavored conformal theory. Just as in the chiral setup \cite{massivekw}, $|V(L)|$ decreases with $N_f$ and increases with $m_q$. This is the expected effect of the screening of the color charges due to the sea quarks: the more the theory is ``unquenched'' (large $N_f$, small $m_q$) the more the modulus of the quark-antiquark force is reduced. Finally, the large $L$ behavior of $V(L)$ is not strongly affected by the choice of the cutoff $M_Q$; in the small $L$ region, instead, we see that $|V(L)|$ is an increasing function of $M_Q$. 

The above analysis refers to the {\it connected} part of the potential, i.e. that which is obtained by artificially turning off the mixing of the $\bar Q Q$ state with $\bar Q q, \bar q Q$ pairs which can originate from the decay of the initial meson. The string setup we have constructed is well aware that this decay can happen as the dynamical $\bar q, q$ pair is popped out of the vacuum. Indeed, due to the presence of the dynamical flavor branes, the open string attached to the D7 probe can break, causing the decay. The reason why the connected part of the potential is an interesting object in this and related setups is the fact that, due to the smearing over the large transverse volume, each decay of this kind is suppressed as $1/N_c$ and not as $N_f/N_c$ as it would happen if the flavor branes would be on top of each other. 
Hence the $\bar Q Q$ state is {\it metastable}.

In the present setup the smallest heavy-light mesons which can be produced are nearly massless (due to the smearing, there will always exist a dynamical flavor brane intersecting the probe $Q$ one). The critical length for the decay to happen can be thus estimated as the solution of $2M_Q + V(L_s)\sim 0$. It is possible to verify that, just as in the chiral case \cite{massivekw}, the screening length $L_s$ decreases with $N_f$ and $M_Q$ and increases with $m_q$.

\begin{acknowledgement}
We thank Carlos N\'u\~nez for comments. Work supported by the EU contract MRTN-CT-2004-
005104, the FWO - Vlaanderen project G.0235.05, the
Federal Office for STC Affairs programme P6/11-P, the
Belgian FRFC (G.2.4655.07), IISN (G.4.4505.86), the
IAPP (Belgian Science Policy), a NWO VIDI grant
016.069.313 and INTAS contract 03-51-6346, the MEC and  FEDER (G.
FPA2005-00188), the Spanish Consolider-Ingenio 2010 Programme CPAN
(CSD2007-00042) and Xunta de Galicia (Conselleria de Educacion and
G.PGIDIT06PXIB206185PR).

F. B. and A. L. C. would like to thank the italian students, parents and scientists for 
their activity in support of public education and research.

\end{acknowledgement}


\end{document}